# Phase behaviors of binary mixtures composed of banana-shaped and calamitic mesogens

M. Cvetinov[a]†, D. Ž. Obadović[a], M. Stojanović[a], A. Vajda[b], K. Fodor-Csorba[b],
N. Eber[b], and I. Ristić[c]

[a] Department of Physics, University of Novi Sad, Novi Sad, Serbia
[b] Wigner Research Center, Institute for Solid State Physics and Optics,
Hungarian Academy of Sciences, Budapest, Hungary
[c] Technological Faculty, University of Novi Sad, Novi Sad, Serbia

In this work, five mixtures with different concentrations of banana-shaped and calamitic compounds have been prepared and subsequently studied by polarizing optical microscopy, differential scanning calorimetry, and X-ray diffraction on non-oriented samples. The phase sequences and molecular parameters of the binary systems are presented.

## 1. Introduction

Calamitic compounds are prototypal liquid crystals and their nematic mesophase profoundly influenced the industry of liquid crystalline displays. Miscibility studies of calamitic compounds were common during the last thirty years.[1] Banana-shaped compounds represent a new class of thermotropic liquid crystals with a nonconventional architecture and an ability to exhibit mesomorphic properties (banana phases B1–B8) different from those of classical liquid crystals.[2,3] Nematic phases are rare in banana-shaped compounds with small molecular bend angles.[4] Yet, induction of nematic phase has been achieved by mixing banana-shaped and calamitic compounds.[5]

Molecular design and synthesis are viable but expensive ways to influence the transition temperatures and properties of mesophases. The needed properties can rather be reached by mixing compounds with various molecular shapes and properties, than by looking for a pure compound with definite properties.[6,7] Miscibility studies with bent-core compounds are still in their infancy.[8–13] Following this reasoning, we have studied binary mixtures of a bent-core and a calamitic nematic liquid crystal.

One mixing component was the banana-shaped nematogenic compound 4,6-dichloro-1,3-phenylene bis [4'-(10-undecen-1-yloxy)-1,1'-biphenyl-4-carboxylate] (I).[14] The calamitic compound was the 4-n-decyloxy-benzoicacid-[4-ndodecyloxy-phenylester] (II), which exhibits the smectic SmA phase in a very narrow temperature range of 2 °C and the SmC phase that pans a temperature range of 11.5 °C.[15] Five mixtures with different concentrations have been prepared and subsequently studied by polarizing optical microscopy, differential scanning calorimetry, and X-ray diffraction on non-oriented samples. Semi-empirical quantum chemical calculations were performed as well. The chemical structures of the studied compounds are depicted in Fig. 1.

* Project supported by the Ministry of Education and Science of the Republic of Serbia (Grant No. OI171015), the Hungarian Research Fund OTKA K81250, and the SASA-HAS Bilateral Scientific Exchange Project #9.
†Corresponding author. E-mail: cvelee@gmail.com

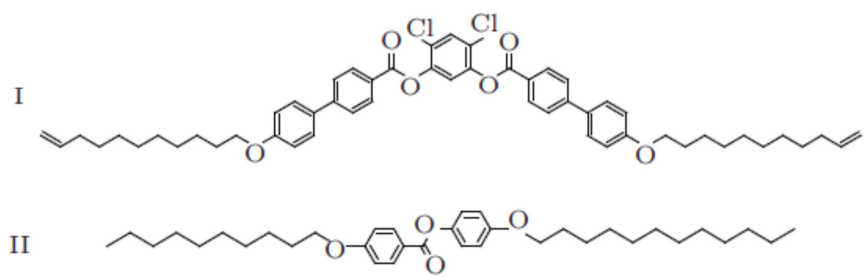

Fig. 1. Chemical structures of the compounds under study.

## 2. Experimental

Sequences of phases and phase-transition temperatures were determined from the characteristic textures and their changes observed in a polarizing optical microscope (POM) Carl Zeiss Jena on non-oriented samples. A hot-stage with platinum–rhodium thermocouple was employed for controlled heating and cooling of the sample. The cell thickness of 12.5 μm was achieved using a mylar spacer that separated the glass substrates. Clamped together, the cell was filled by capillary action with the studied sample.

Thermal properties of the samples were measured using a DuPont DSC 910. Hermetically sealed aluminium pans containing 1–3 mg of the sample were prepared. Calibration of the equipment was carried out by indium as the temperature standard. All samples were melted at 120 °C, to ensure complete melting and resetting the thermal history, and cooled at room temperature for two days. The non-isothermal analysis was performed from 20 °C to 150 °C with heating rate 10 °C·min$^{-1}$. Subsequently, the sample was allowed to cool spontaneously to room temperature, thus only $T_{I-N}$ phase temperatures could be determined unambiguously from DSC in cooling.

In order to obtain more structural information and with the intention of affirming or refuting previously reported sequences of phases, non-oriented samples were investigated by X-ray diffraction in Brag–Brentano $\theta : 2\theta$ geometry using a conventional powder diffractometer, Seifert V-14, equipped with an automated high-temperature kit Paar HTK-10. Diffraction was detected at the Cu K$\alpha$ radiation of 0.154059 nm, with a Ni filter installed and in absence of a monochromator. Calibration was performed employing the two most intense platinum lines. Continual scanning was employed with a scanning speed of 1° $2\theta$min$^{-1}$.

## 3. Results and discussion

The goal of the present study was to test the miscibility of the bent-core compound I with the rod-like material II, and to study the mesomorphic behavior of their binary mixtures. For the detailed study, five mixtures, M1 to M5, have been prepared with 20, 40, 50, 60, and 80 wt% of the bent-core component I, respectively. The phase transition temperatures of pure compounds and their mixtures in heating and cooling were detected by differential scanning microscopy and polarizing optical microscopy, which are presented in Tables 1 and 2. The identification of the phases (indicated in Table 1) was made by POM studies.

Table 1. Sequence of phases and phase transition temperatures *T* end enthalpies *ΔH*. All these data were obtained by DSC in heating. (I = isotropic, N = Nematic, A = Smectic A, C = Smectic C, $C_n$ = $n^{th}$ crystalline phase, • the phase exists).

| Mix | $C_1$ | *T*/°C | *ΔH*/J·g$^{-1}$ | $C_2$ | *T*/°C | *ΔH*/J·g$^{-1}$ | C | *T*/°C | *ΔH*/J·g$^{-1}$ | A | *T*/°C | *ΔH*/J·g$^{-1}$ | N | *T*/°C | *ΔH*/J·g$^{-1}$ | I |
|---|---|---|---|---|---|---|---|---|---|---|---|---|---|---|---|---|
| II | • | 60.5 | 12.5 | • | 68.5 | 99.0 | • | 79.0 | * | • | | | | 81.0 | 11.6 | • |
| M1 | • | 57.0 | 41.7 | • | 67.0 | 120.8 | | | | | | | • | 80.0 | 10.9 | • |
| M2 | • | 58.0 | 19.5 | • | 61.0 | 54.0 | | | | | | | • | 79.0 | 3.0 | • |
| M3 | • | 49.5 | 6.3 | • | 60.0 | 119.4 | | | | | | | • | 82.0 | 5.4 | • |
| M4 | • | 49.5 | 12.5 | • | 60.5 | 136.3 | | | | | | | • | 84.0 | 3.8 | • |
| M5 | • | 59.0 | 96.0 | • | 69.5 | 42.0 | | | | | | | • | 89.0 | 7.5 | • |
| I | • | 77.5 | 58.7 | | | | | | | | | | • | 104.9 | 1.4 | • |

* Enthalpies are dependent on the sensitivity of the device. Reliable data are not given.

Table 2. Sequence of phases and phase transition temperatures *T* obtained by DSC and POM in cooling.

| Mix | $C_1$ | *T*/°C | C | *T*/°C | A | *T*/°C | N | *T*/°C | I |
|---|---|---|---|---|---|---|---|---|---|
| II | • | 66.5 | • | 78.0 | • | | | 80.0 | • |
| M1 | • | 60.5 | | | | | • | 78.5 | • |
| M2 | • | 58.0 | | | | | • | 76.0 | • |
| M3 | • | 57.5 | | | | | • | 80.5 | • |
| M4 | • | 57.0 | | | | | • | 82.0 | • |
| M5 | • | 58.0 | | | | | • | 89.0 | • |
| I | • | 60.9 | | | | | • | 102.7 | • |

DSC thermograms in heating and cooling are presented on Figs. 2 and 3 respectively. Phase transition temperatures, which were determined by DSC and/or POM, are marked with arrows. Both melting and clearing point temperatures were lowered in mixtures with respect to pure compounds.

In the compound II, the SmA–SmC phase transition is a second-order phase transition that is characterized by a very slight shift in the baseline of the thermogram, indicating slight but abrupt change in heat capacity.[16] The distinguishing characteristic of first-order phase transitions is the discontinuity in entropy and enthalpy. A general trend that the melting enthalpy changes are an order of magnitude larger than the clearing ones is observed.[17] Due to the bent molecular shape, the enthalpy change at N–I phase transition of pure compound I is lowered with respect to classic calamitic compounds.[18]

Banana-shaped molecules, due to their rigidity and bent molecular shape, obstruct molecular ordering in layers, ultimately causing the disappearance of smectic phases in all mixtures. Hence, polymorphism of starting compound II is not inherited in mixtures. A nematic phase, which was virtually nonexistent in compound II, emerged in the mixture with the lowest concentration of compound I (M1) and remained stable throughout the whole concentration range. Moreover, the nematic temperature range expanded with the increase of concentration of banana-shaped compound I, reaching its widest extent of 31 °C in M5.

Microphotographs of some characteristic textures of the various mesophases obtained in cooling of non-homogenously aligned planar samples are presented in Fig. 4. Textures obtained in heating are similar to textures obtained in cooling.

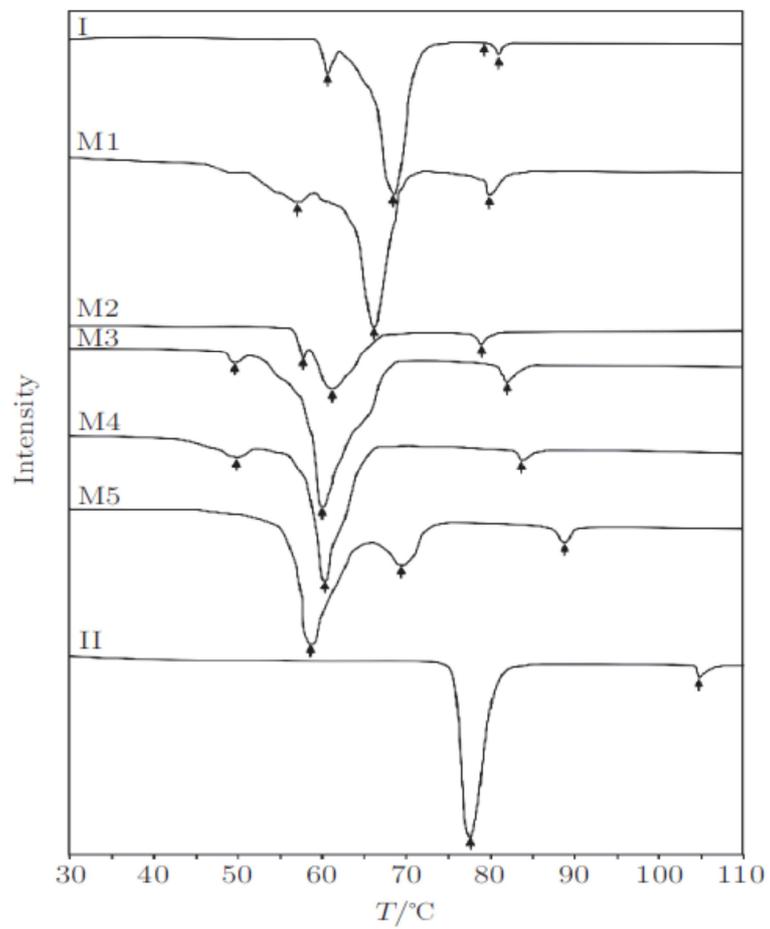

Fig. 2. DSC thermograms of pure compounds and mixtures in heating.

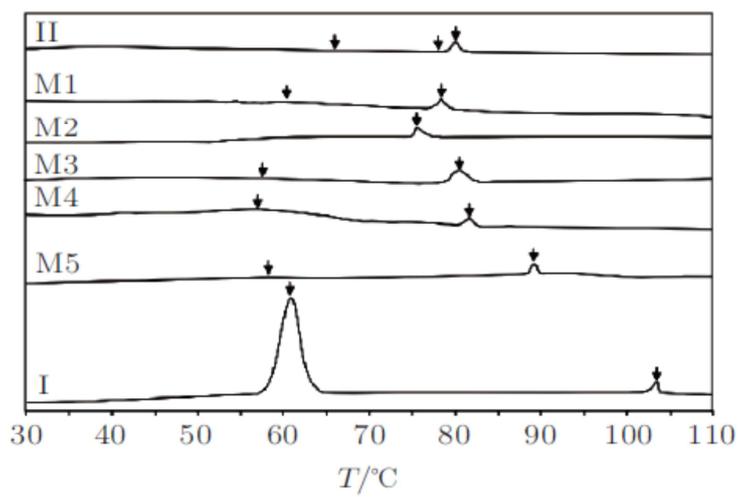

Fig. 3. DSC thermograms of pure compounds and mixtures in cooling.

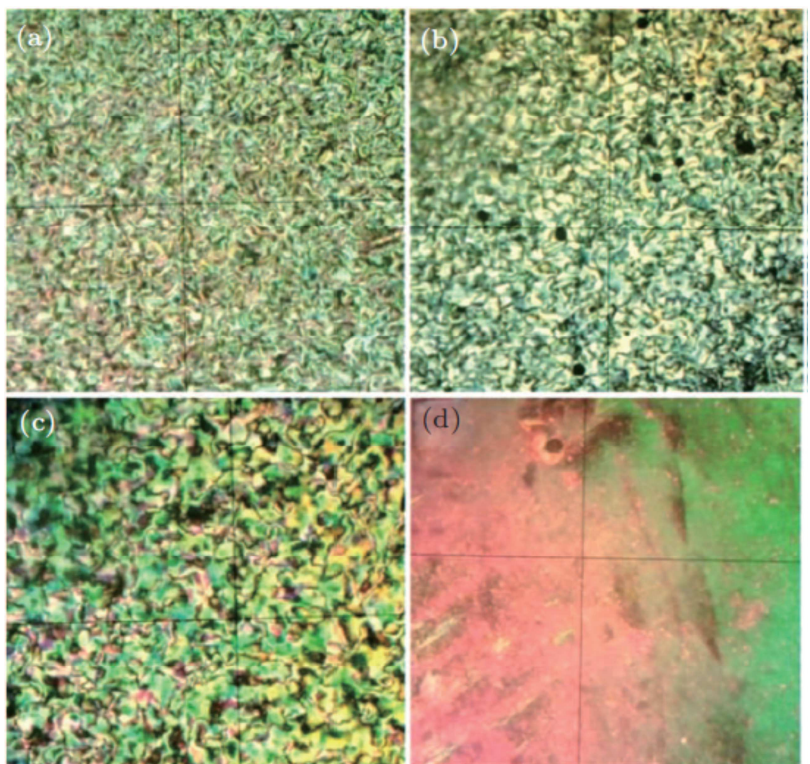

Fig. 4. (color online) Micrographs of nematic phases of non-oriented samples of: (a) M3 at 75 °C, (b) M4 at 80 °C, (c) M5 at 85 °C, and (d) I at 88 °C, in cooling. All micrographs are 300 μm wide.

A thread-like texture is predominant in the nematic phase of mixtures (Figs. 4(a) and 4(b)), where s = ±1/2 point defects are clearly visible on bigger threads of M4. The marbled texture occurs in the nematic phase of M5 (Fig. 4(c)), showing some threads across the whole viewing area. Marbled texture is characteristic of banana shaped compound I (Fig. 4(d)), where due to a change in birefringence from in-plane and outof- plane variations of the director, a slight change in color is observed.[19] Because of its predominance in banana nematic phases, the inheritance of the marbled texture in mixtures with the highest content of compound I is expected, as shown in Fig. 4(c).

Figures 5–7 show diffractograms of starting compounds (I and II) and M1. Molecular parameters were calculated according to Bragg's law $n\lambda = 2d \sin\theta$, where $n$ is the order of reflection, $\lambda$ is the wavelength of incident radiation, $\theta$ is the Bragg's scattering angle, and $d$ is the repetitive distance which is to be determined.[20,21]

Molecular parameters are presented in Table 3. If existent, second-order reflection peaks are not shown in Table 3. Data are adjusted for zero shift and sample displacement errors.

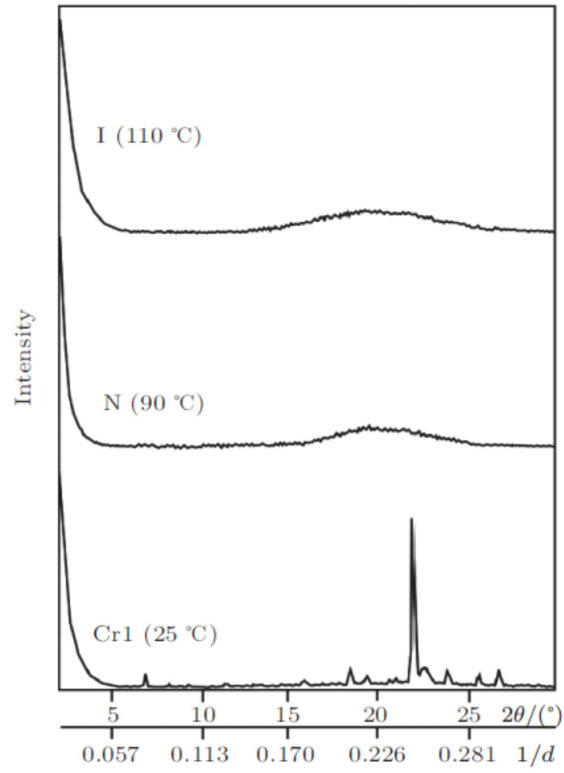

Fig. 5. X-ray diffractogram of I.

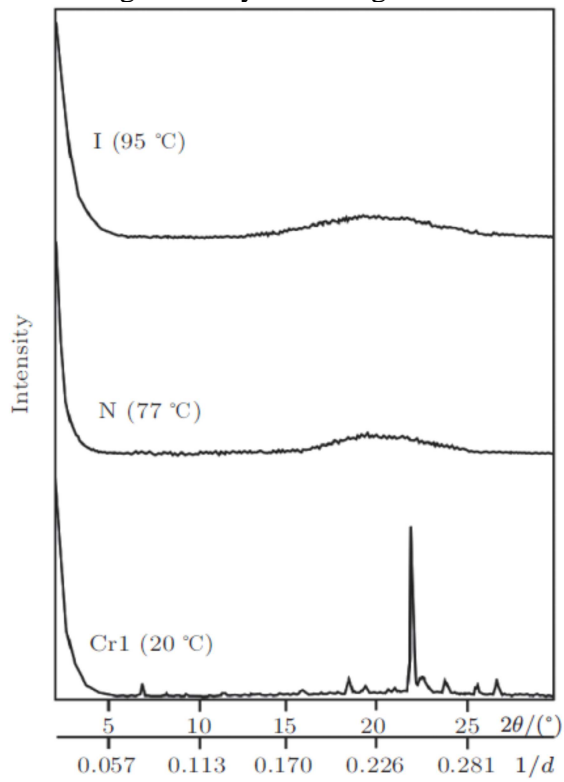

Fig. 6. X-ray diffractogram of M1.

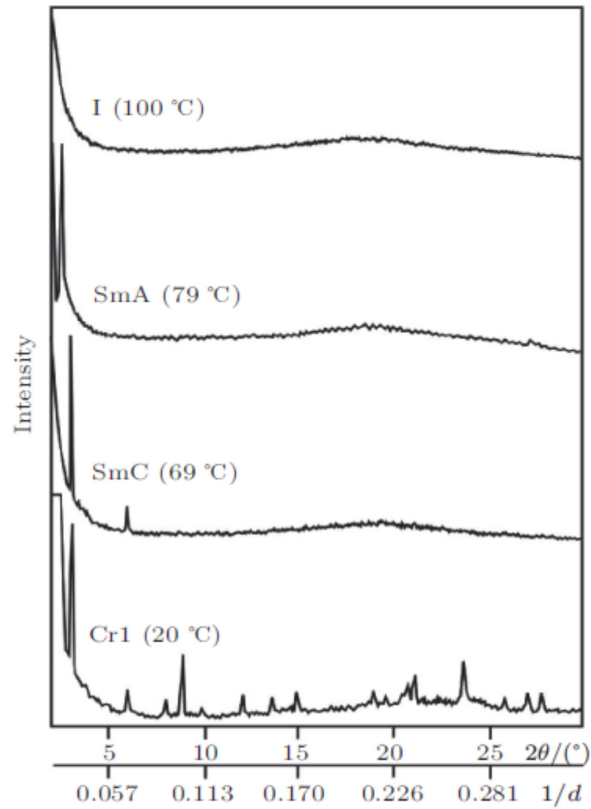

Fig. 7. X-ray diffractogram of II.

Table 3. Molecular parameters of the investigated mixtures for all observed phases at a fixed temperature $T$: angles corresponding to the reflection peaks $2\theta$, effective layer thickness $d$ (in Å ; the error of measurements was $\sigma_d \approx 0.01$ Å ), average repeat distance $D$ (in Å ; the error of measurements was $\sigma_D \approx 0.02$ Å ).

| Mix | $T/°C$ | $2\theta/(°)$ | $d/$ Å | $D/$ Å |
|---|---|---|---|---|
| II | 100 (I) | 19.0 | | 4.67 |
| | 79 (SmA) | 2.40   19.6 | 36.7 | 4.52 |
| | 69 (SmC) | 3.00   19.8 | 29.4 | 4.48 |
| M1 | 95 (I) | 18.7 | | 4.74 |
| | 77 (N) | 19.3 | | 4.59 |
| M2 | 90 (I) | 17.5 | | 5.06 |
| | 75 (N) | 18.3 | | 4.84 |
| M3 | 90 (I) | 18.0 | | 4.92 |
| | 70 (N) | 18.8 | | 4.71 |
| M4 | 86 (I) | 18.2 | | 4.87 |
| | 61 (N) | 18.9 | | 4.69 |
| M5 | 103 (I) | 18.8 | | 4.71 |
| | 79 (N) | 19.5 | | 4.54 |
| I | 110 (I) | 19.3 | | 4.59 |
| | 90 (N) | 20.0 | | 4.44 |

The broad and intense diffuse amorphous halo observed in the crystalline phase of pure compound II is indicative of the poor crystallinity of the sample (Fig. 7).[22] Yet, in the crystalline phase of pure compound I and even in the mixtures with the lowest content of compound I (M1, Fig. 6), good crystallinity is observed, evident by the almost nonexistent amorphous halo and the existence of numerous sharp peaks at wide Bragg angles.

The low angle reflection peak at $2\theta = 2.40°$ in the SmA phase of pure compound II corresponds to a smectic layer thickness of 36.78 Å. SmA–SmC is a second-order phase transition characterized by the gradual change of the molecular tilt angle.[23] In the SmC phase, molecules attained a tilt angle of 40.5° with respect to layer normal, indicated by a low angle diffraction peak at $2\theta = 3.00°$.

In addition to the isotropic phase, all mesophases are further characterized by a broad diffusion peak which spans in the range of $2\theta = 12°–26°$ with its maximum appearing at $2\theta = 18°–20°$, corresponding to the lateral distance between the long molecular axes. Analysis of the diffraction profiles has shown that the center of the broad diffraction peak shifts slightly toward larger angles with decreasing the temperature.

This indicates that the average intermolecular distance decreases during the successive phase transitions on cooling, i.e., the packing becomes slightly denser. This is due to the fact that supramolecular stacking is governed by non-covalent intermolecular interactions (hydrogen bonding and dipole and aromatic interaction) which have energies only a few times higher than the thermal energy (kT).[24,25]

Molecular models were constructed to give us an insight into the self-assembly of the molecules in the mesophase. Quantum chemistry calculation, such as the density function and the discrete variational method, has been well used in molecular structural optimization and the characterization of materials. Owing to their fast computational time, the semiempirical methods are frequently employed to evaluate the molecular parameters of liquid crystalline substances.[26,27]

Among the tested semi-empirical methods, RM1 yielded the least average errors for bond lengths while retaining low average errors for bond angles.[28] Hence, RM1 is the most adequate semi-empirical method for determination of the length of whole molecules and it was used in this research because of the large number of non-hydrogen atoms in the molecules studied. Subsequently, the results of computation were compared with the X-ray measurements.

The electrostatic potential map of compound II is shown in Fig. 8. The density of electrons appears as colors changing from blue to red; it is the highest in the vicinity of atoms characterized by high electronegativity (principally at oxygen atoms).[29] The dihedral angle between the two neighbouring phenyl rings of the molecule is 44.15°. Recalculation using RM1 parameterization yielded 38.66 Å for the molecular length of compound I, measured between terminal C atoms.

The linear length of its rigid molecular core, measured between its terminal -OCH2- groups, is 13.60 Å. The shorter alkyl chain has a linear length of 11.32 Å while the longer one has a length of 13.83 Å.

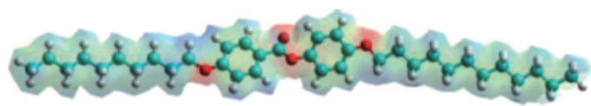

Fig. 8. (color online) Electrostatic potential map of the minimum energy conformation of compound II.

The electrostatic potential map of compound I is shown in Fig. 9. The calculated bending angle between the two arms of the molecule is 121.32°, slightly less than the previously reported value. The angle between the two neighboring phenyl rings in both arms of the molecule is 44.1°. Recalculation using RM1 parameterization yielded 46.5 Å for the molecular length of compound I; the linear length of half of its rigid molecular core is 14.03 Å and the linear length of each alkyl chain is 14.44 Å .

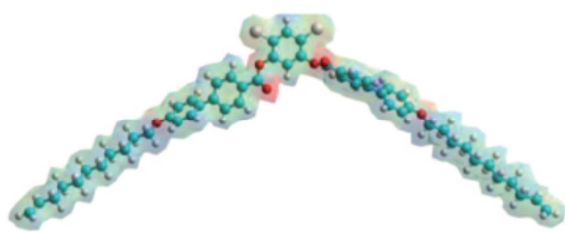

Fig. 9. (color online) Electrostatic potential map of the minimum energy conformation of compound I.

In non-polar mesogens, such as compound II, a slightly smaller value of smectic layer spacing obtained by means of X-ray diffraction than the calculated molecular length is often encountered in the SmA phase, primarily due to the existence of the de Vries SmA phase.[30,31] De Vries proposed that already in the SmA phase, molecules could be tilted, but in unbiased rotation, making the SmA phase optically uniaxial. Another explanation for this discrepancy is that due to complete head-to-tail molecular disorder, intercalation of ends of alkyl chains into neighbouring layers is probable.[32] With intercalation present, there is increased probability of the existence of gauche defects in long alkyl chains, which further shorten the effective molecular length, hence, shortening the smectic interlayer spacing.[33]

Determination of molecular shapes allows the construction of a model of the molecular packaging in the nematic mesophase of mixtures. For molar concentrations of 38.1 mol% of compound I in M3, the proposed molecular packaging is shown in Fig. 10. At low temperatures, flexible alkyl chains assume fully extended conformation. In the nematic phase, rigid banana-shaped molecules hinder the uniform orientation of calamitic molecules. This gives rise to microdomains composed of calamitic molecules oriented parallel to one molecular arm of banana-shaped molecules. Both calamitic and banana-shaped molecules retain their nonpolar order, due to the equal probability of molecules pointing in any of two possible directions (up and down, or left and right).

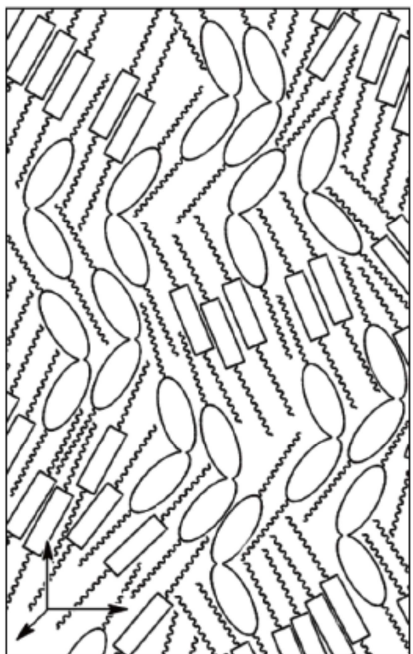

Fig. 10. Proposed molecular packaging of molecules in the nematic phase of M3.

## 4. Conclusions

The nematic phase is observed in all mixtures, reaching its widest temperature range of 31.0 °C at the concentration of 80 wt% of compound I (M5). All mixtures exhibited only slightly lowered crystallization temperatures in respect to the starting compounds. The expected finding is the loss of smectic phases in mixtures, which is caused by the pronounced difference in molecular shapes of starting compounds. Therefore, the full polymorphism of the starting calamitic compound is not preserved in any mixture.